\documentclass[letter,aps,showpacs,superscriptaddress,nofootinbib,tightenlines,twocolumn,prl]{revtex4}

\usepackage{amsfonts}
\usepackage{multirow}
\usepackage{mathrsfs}
\usepackage{graphicx}
\usepackage{amsmath}
\usepackage{amssymb}
\usepackage{bm}
\usepackage{bbm}
\usepackage{color}

\def\OMIT#1{}

\newcommand{\nn}{\nonumber}

\newcommand{\beq}{\begin{equation}}
\newcommand{\eeq}{\end{equation}}
\newcommand{\bqa}{\begin{eqnarray}}
\newcommand{\eqa}{\end{eqnarray}}

\begin{document}
\title{\mbox{}\\[10pt]
Next-to-next-to-leading-order QCD corrections to hadronic width of pseudoscalar quarkonium
}

\author{Feng Feng\footnote{F.Feng@outlook.com}}
\affiliation{Institute of High Energy Physics and Theoretical Physics Center for
Science Facilities, Chinese Academy of
Sciences, Beijing 100049, China\vspace{0.2cm}}
\affiliation{China University of Mining and Technology, Beijing 100083, China\vspace{0.2cm}}

\author{Yu Jia\footnote{jiay@ihep.ac.cn}}
\affiliation{Institute of High Energy Physics and Theoretical Physics Center for
Science Facilities, Chinese Academy of
Sciences, Beijing 100049, China\vspace{0.2cm}}
\affiliation{School of Physics, University of Chinese Academy of Sciences,
Beijing 100049, China\vspace{0.2cm}}
\affiliation{Center
for High Energy Physics, Peking University, Beijing 100871,
China\vspace{0.2cm}}

\author{Wen-Long Sang\footnote{wlsang@ihep.ac.cn}}
 \affiliation{School of Physical Science and Technology, Southwest University, Chongqing 400700, China\vspace{0.2cm}}

\date{\today}
\begin{abstract}
We compute the next-to-next-to-leading order (NNLO) QCD corrections to the hadronic decay rates of the
pseudoscalar quarkonia, at the lowest order in velocity expansion.
The validity of NRQCD factorization for inclusive quarkonium decay process, for the first time,
is verified to relative order $\alpha_s^2$.
As a byproduct, the renormalization group equation (RGE) of the leading NRQCD 4-fermion operator
${\cal O}_1({}^1S_0)$ is also deduced to this perturbative order.
By incorporating this new piece of correction together with available relativistic corrections,
we find that there exists severe tension between the state-of-the-art NRQCD predictions
and the measured $\eta_c$ hadronic width, and in particular the branching fraction of $\eta_c\to \gamma\gamma$.
NRQCD appears to be capable of accounting for $\eta_b$ hadronic decay to a satisfactory degree,
and our most refined prediction is ${\rm Br}(\eta_b\to \gamma\gamma) = (4.8\pm 0.7)\times 10^{-5}$.
\end{abstract}



\maketitle

Heavy quarkonium decay has historically played a preeminent role in establishing asymptotic freedom of QCD~\cite{Appelquist:1974zd,DeRujula:1974rkb}.
Due to the nonrelativistic nature of heavy quark inside a quarkonium,
the decay rates are traditionally expressed as the squared bound-state wave function
at the origin multiplying the short-distance quark-antiquark annihilation decay rates.
With the advent of the modern effective-field-theory approach,
the nonrelativistic QCD (NRQCD),
this factorization picture has been put on a firmer ground, and one is allowed to systematically
include the QCD radiative and relativistic corrections when tackling various quarkonium
decay and production processes~\cite{Bodwin:1994jh}.

The aim of this Letter is to critically scrutinize one of the
most basic quantities in
the area of quarkonium physics, {\it i.e.}, the hadronic widths of ${}^1S_0$ charmonia and bottomonia.
The latest Particle Data Group (PDG) compilation lists the total widths
$\Gamma_{\rm had}(\eta_c)=31.8\pm 0.8$ MeV, and
$\Gamma_{\rm had}(\eta_b)= 10^{+5}_{-4}$ MeV~\cite{Olive:2016xmw}.
It is rather challenging, if not impossible, for lattice QCD and other influential
nonperturbative methods to accurately account for these hadronic decay widths.
However, these simple yet important observables naturally constitute the ideal
candidates to critically examine the validity of the NRQCD factorization approach.

According to NRQCD factorization~\cite{Bodwin:1994jh}, through the relative order $v^2$,
the inclusive hadronic decay rate of the pseudoscalar quarkonium, say, $\eta_c$,
can be written as
\bqa
&&\Gamma(\eta_c \to {\rm LH}) =  { F_1({}^1S_0) \over m^2} \langle \eta_c| {\cal O}_1({}^1S_0) |\eta_c\rangle
\nn
\\
&& +
{G_1({}^1S_0)\over m^4}
\langle \eta_c| {\cal P}_1({}^1S_0) |\eta_c\rangle  + {\cal O}(v^3 \Gamma),
\label{etac:decay:NRQCD:factorization}
\eqa
where ${\cal O}_1({}^1S_0)=\psi^\dagger \chi \chi^\dagger \psi$,
${\cal P}_1({}^1S_0)=\frac{1}{2}\big[\psi^\dagger\chi \chi^\dagger
(-\tfrac{i}{2}\tensor{\mathbf{D}})^{2}\psi +{\rm h.c.} \big]$.
Here $\psi,\,\chi$ represent the quark and anti-quark Pauli spinor fields in NRQCD,
and $\mathbf{D}$ denotes the spatial part of the gauge covariant derivative.
In Refs.~\cite{Bodwin:2002hg,Brambilla:2008zg},
more complete NRQCD factorization formulae are presented through the relative order $v^4$.
Since the explosion of the number of poorly-constrained operator matrix elements
severely hampers the predictive power of NRQCD,
in this Letter we will be contented with the accuracy of the velocity expansion
as prescribed in (\ref{etac:decay:NRQCD:factorization}). Some crude power-counting argument
estimates that those neglected terms in (\ref{etac:decay:NRQCD:factorization})
may yield a contribution as large as 25\%~\cite{Bodwin:2002hg}.

It is convenient to organize these short-distance coefficients in terms of
perturbative series expansion:
\begin{subequations}
\bqa
&& F_1({}^1S_0)= \frac{\pi C_F \alpha_s^2}{N_c} \left\{ 1+ {\alpha_s\over \pi} f_1+
{\alpha_s^2\over \pi^2} f_2+\cdots\right\},
\label{parametrization:F1}
\\
&& G_1({}^1S_0)= -{4\pi C_F \alpha_s^2 \over 3 N_c} \left\{ 1+ {\alpha_s\over \pi} g_1+
\cdots\right\}.
\eqa
\label{parametrization:F1:G1}
\end{subequations}
The ${\cal O}(\alpha_s)$ correction to the short-distance
coefficient $F_1({}^1S_0)$ was first computed in Refs.~\cite{Barbieri:1979be,Hagiwara:1980nv}.
The tree-level contribution to $G_1({}^1S_0)$ was first given in Refs.~\cite{Keung:1982jb,Bodwin:1994jh}.
The ${\cal O}(\alpha_s)$ correction to $G_1({}^1S_0)$ was recently calculated in
Ref.~\cite{Guo:2011tz}.
As a crosscheck, we recalculate these ${\cal O}(\alpha_s)$ corrections and find
\begin{subequations}
\bqa
 f_1 &=& \frac{\beta_0}{2}\ln\frac{\mu_R^2}{4m^2} +
\bigg(\frac{\pi^2}{4}-5\bigg)C_F+\bigg(\frac{199}{18}-\frac{13\pi^2}{24}\bigg)C_A
\nn\\
&& -\frac{8}{9}n_L -\frac{2 n_H}{3}\ln2,
\label{f1:expressions}
\\
 g_1 &=& \frac{\beta_0}{2}\ln\frac{\mu_R^2}{4m^2} - C_F\ln\frac{\mu_\Lambda^2}{m^2}
 -\bigg(\frac{49}{12}-\frac{5\pi^2}{16}-2\ln2\bigg)C_F
\nn\\
&&+\bigg(\frac{479}{36}-\frac{11\pi^2}{16}\bigg)C_A -\frac{41}{36}n_L - \frac{2 n_H}{3}\ln2.
\label{g1:expressions}
\eqa
\label{f1:g1:expressions}
\end{subequations}
$\beta_0 = {11\over 3}C_A - {4\over 3} T_F n_f$ is
the one-loop coefficient of the QCD $\beta$-function, where $T_F={1\over 2}$,
and $n_f$ signifies the number of active
quark flavors. In this Letter, we choose to include the heavy quark as the ``active" flavor, {\it i.e.},
take $n_f=n_L+n_H$, where $n_L$ labels the number of light quark flavors
($n_L=3$ for $\eta_c$, $4$ for $\eta_b$), and $n_H=1$ is the number of heavy quark.
The $SU(N_c)$ Casimirs $C_F={N_c^2-1\over 2N_c}$, $C_A=N_c$,
where we will eventually take the number of colors $N_c=3$.
The occurrences of the $\beta_0\ln \mu_R$ term in (\ref{f1:g1:expressions})
are constrained by the independence of the decay rate on the
renormalization scale $\mu_R$.
The emergence of the factorization scale $\mu_\Lambda$ in (\ref{g1:expressions})
reflects that the NRQCD 4-fermion operator ${\cal O}_1({}^1S_0)$
depends on the renormalization point $\mu_\Lambda$ such as to ensure the
$\mu_\Lambda$-independence of the decay rate.

If setting $n_H=0$ in (\ref{f1:g1:expressions}), as was commonly practiced
in the preceding perturbative calculations, our NLO short-distance coefficients
$f_1$ and $g_1$ will reproduce the values reported in
Refs.~\cite{Barbieri:1979be,Hagiwara:1980nv} and \cite{Guo:2011tz}.

The goal of this Letter is to compute the NNLO perturbative coefficient $f_2$ in (\ref{parametrization:F1}).
To date, perturbative calculations beyond NLO have been conducted only for a few {\it exclusive} processes
involving quarkonium, exemplified by ${\cal O}(\alpha_s^2)$ corrections to $\Upsilon(J/\psi)\to e^+e^-$~\cite{Czarnecki:1997vz,Beneke:1997jm} (Notice the ${\cal O}(\alpha_s^3)$ coefficients were also available recently~\cite{Marquard:2014pea,Beneke:2014qea}),
$\eta_{b,c}\to\gamma\gamma$~\cite{Czarnecki:2001zc,Feng:2015uha}, $\chi_{c0,2}\to\gamma\gamma$~\cite{Sang:2015uxg},
and $B_c\to \ell \nu$~\cite{Onishchenko:2003ui,Chen:2015csa}, as well as the ${\cal O}(\alpha_s^2)$ correction
to the $\gamma\gamma^*\to \eta_{c,b}$ transition form factor~\cite{Feng:2015uha}.
Only two-loop virtual corrections are required in calculating these hard matching coefficients,
since they correspond to exclusive quarkonium decays or productions.
In contrast, in order to compute $f_2$ to NNLO in $\alpha_s$, one must incorporate both real as well as
virtual corrections, which turns out to be much more demanding than the aforementioned work.

\begin{figure}[tb]
\centering
\includegraphics[width=0.5\textwidth]{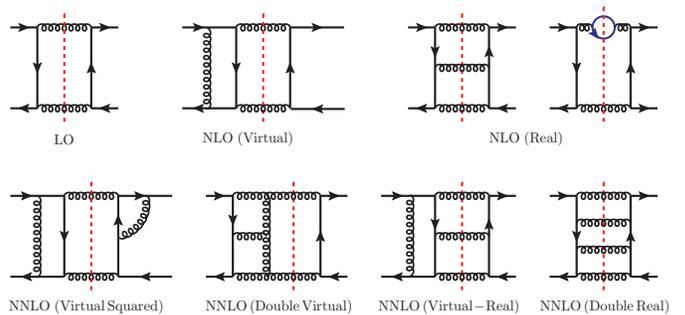}
\caption{Representative cut Feynman diagrams responsible for the quark reaction
$c\bar{c}({}^1S^{(1)}_0)\to c\bar{c}({}^1S^{(1)}_0)$
through NNLO in $\alpha_s$. The vertical dashed line denotes the
Cutkosky cut. \label{Cut:Feynman:Diagrams}}
\end{figure}

This Letter reports the very first effort to compute the full NNLO corrections to the
{\it inclusive} hadronic decay of heavy quarkonium.
To determine the short-distance coefficients via perturbative matching procedure,
it is most convenient to appeal to the optical theorem, to start from the
forward-scattering quark amplitude for $c\bar{c}({}^1S^{(1)}_0)\to c\bar{c}({}^1S^{(1)}_0)$,
then extract the respective imaginary part by invoking the Cutkosky rule.
Some typical cut Feynman diagrams for such a quark-level process through three-loop order are illustrated in Fig.~\ref{Cut:Feynman:Diagrams}.
In passing, it might be worth mentioning that, the parton-level calculation
considered here somewhat resembles the NNLO correction to $gg\to t\bar{t}$~\cite{Czakon:2013goa},
yet exactly sitting at the $t\bar{t}$ threshold.
Moreover, it is also convenient to use the covariant trace technique to expedite the projection of the
$c\bar{c}$ pairs onto the spin-singlet states. Prior to performing the loop integration,
we neglect the relative momentum between $c$ and $\bar{c}$ in both initial and final states,
which amounts to enforcing the $c\bar{c}$ in the $S$-wave state,
and allows us to directly extract the short-distance coefficients at $v^0$ accuracy.
Dimensional regularization (DR), with the spacetime dimensions $D=4-2\epsilon$,
is utilized to regularize both UV and IR divergences.

We use the packages {\tt QGraf}~\cite{Nogueira:1991ex} and {\tt FeynArts}~\cite{Hahn:2000kx} to
generate the
three-loop Feynman diagrams and corresponding forward-scattering amplitudes in Feynman gauge.
Roughly 1700 diagrams are generated.
We use {\tt FeynCalc/FormLink}\cite{Mertig:1990an,Feng:2012tk} to conduct the Dirac/color trace operations.
After imposing the Cutkosky rule, all the cut Feynman diagrams can be divided into four topologies, which
are dubbed as ``Virtual Squared", ``Double Virtual",  ``Virtual-Real", and ``Double Real", respectively,
as can be visualized in Fig.~\ref{Cut:Feynman:Diagrams}.
The first class involves the squared one-loop amplitude for $c\bar{c}({}^1S^{(1)}_0)\to gg$,
which can be readily obtained analytically.

The pressing challenge is how to accurately conduct the multi-body phase space integration in DR,
especially for the ``Virtual-Real" and ``Double Real" types, which are plagued with
severe IR-divergences descending from order-$\epsilon^{-4}$.
In principle, one may invoke some sophisticated IR-divergence subtraction schemes
that are widely employed for the NNLO calculation in hadron colliders.
Fortunately, since we are only interested in the inclusive annihilation decay rate, it is much
more efficient to follow a powerful trick, which was first introduced to expedite calculating the
NNLO correction to inclusive Higgs hadroproduction rate~\cite{Anastasiou:2002yz}.
The key idea is to convert a phase-space integral into a loop integral, which is facilitated by
the following simple identity for the $i$-th cut propagator~\cite{Anastasiou:2002yz,Gehrmann-DeRidder:2003pne}:
\begin{equation}
\int\frac{d^D p_i}{(2\pi)^{D}} 2 \pi\,i\,\delta_+(p_i^2) = \int\frac{d^D p_i}{(2\pi)^{D}} \Bigg(\frac{1}{p_i^2 + i\varepsilon} - \frac{1}{p_i^2 - i\varepsilon}\Bigg).
\label{cut:trick}
\end{equation}
Since the differentiation operation involved in the integration-by-parts (IBP) identities
are insensitive to the $i\varepsilon$, one can apply the IBP method to phase-space integration just as in loop integration~\cite{Anastasiou:2002yz}. Therefore, for the ``Virtual-Real" and ``Double Real"-type diagrams,
we can also utilize the packages {\tt Apart}~\cite{Feng:2012iq} and {\tt FIRE}~\cite{Smirnov:2014hma} to
conduct partial fraction and the corresponding IBP reduction.
Finally, we end up with 93 MIs for the ``Double Virtual" type of diagrams,
89 MIs for the ``Virtual-Real" type of diagrams, and 32 MIs for ``Double Real" type of diagrams,
respectively. To our knowledge, this work represents the first application of the trick (\ref{cut:trick}) in
higher-order calculation involving quarkonium.

We then use {\tt FIESTA}~\cite{Smirnov:2013eza} to perform sector decomposition for all the MIs.
For each decomposed sector, we first use {\tt CubPack}~\cite{CubPack} to conduct the first-round
rough numerical integration.
For those integrals with large estimated errors, two Message-Passing Interface~\cite{MPI}-based
parallelized packages: {\tt PVegas}~\cite{pvegasPaper} and {\tt ParInt}~\cite{parint} are utilized to repeat the numerical integration. For some rather difficult integrals, which mainly stem from the ``Double Virtual" and ``Virtual-Real" sectors,
we have to distribute ${\cal O}(10^{12})$ sample points in order to achieve a tolerable accuracy
for the ${\cal O}(\epsilon^0)$ coefficient.
The numerical integration with such a scale has to be conducted at supercomputer.
One numerically very expensive MI can be analytically extracted from \cite{Bonciani:2009nb},
where satisfactory agreement is found when compared against our numerical result.
The package {\tt Cuba}~\cite{Hahn:2004fe} is also used for cross-checking
most of the integrals.
Roughly speaking, the computational expense of this work is about ${\cal O}(10^5)$ CPU core-hour.

In implementing the renormalization program, we take the ${\cal O}(\alpha_s^2)$ expressions
for $Z_2$, $Z_m$ and $Z_3$
from Refs.~\cite{Broadhurst:1991fy, Melnikov:2000zc,Baernreuther:2013caa},
and renormalize the strong coupling constant to two-loop order in the $\overline{\rm MS}$ scheme.
After the removal of UV divergences,
we obtain the imaginary part of the renormalized $c\bar{c}({}^1S^{(1)}_0)\to c\bar{c}({}^1S^{(1)}_0)$ amplitude
through NNLO in $\alpha_s$. Although each of the four cut topologies contains IR divergences as severe
as $\epsilon^{-4}$, miraculously, only a single IR pole survives in their sum.
Intriguingly, the coefficient of the single pole, to an exquisitely high numerical precision,
can be identified with what was encountered in the NNLO correction to
$\Gamma(\eta_c\to\gamma\gamma)$~\cite{Czarnecki:2001zc,Feng:2015uha}.

Following Refs.~\cite{Czarnecki:2001zc,Feng:2015uha}, we factorize this single IR pole into
LO NRQCD decay matrix element under $\overline{\rm MS}$ prescription.
Finally, the desired short-distance coefficient $f_2$ reads:
\bqa
f_2 &=&  \hat{f}_2+ {3\beta_0^2\over 16} \ln^2\frac{\mu_R^2}{4m^2} +
\left( \frac{\beta_1}{8} + \frac{3}{4} \beta_0 \hat{f}_1\right) \ln\frac{\mu_R^2}{4m^2}
\nn\\
&-& \pi^2\left(C_F^2+\frac{C_A C_F}{2}\right) \ln\frac{\mu_\Lambda^2}{m^2},
\label{f2:expressions}
\eqa
where $\hat{f}_1 \equiv f_1\big|_{\mu_R=2m}$ in (\ref{f1:expressions}),
$\beta_1 = \frac{34}{3}C_A^2 -\frac{20}{3}C_A T_F n_f - 4C_F T_F n_f$ is the two-loop coefficient of
the QCD $\beta$ function. Again, the occurrences of $\beta^i_{0,1}\ln^j\mu_R$ ($i,j=1,2$) are constrained
by the $\mu_R$-independence of the decay rate.
The pivotal achievement of this work is the knowledge of the non-logarithmic constant:
\bqa
\hat{f}_2 &=&  -0.799(13)N_c^2 -7.4412(5)n_L N_c -3.6482(2)N_c
\nn \\
&& +0.37581(3)n_L^2 +0.56165(5)n_L +32.131(5)
\nn\\
&& -0.8248(3)\frac{n_L}{N_c} -\frac{0.67105(3)}{N_c} -\frac{9.9475(2)}{N_c^2}.
\eqa
Notice the coefficient of $N_c^2$ bears the largest uncertainty, which originates from
some most difficult integrals.
Concretely, $\hat{f}_2=-50.1(1)$ for $\eta_c$ hadronic decay,
and $-69.5(1)$ for $\eta_b$ decay.
For completeness, here we also enumerate the numerical values of the
non-logarithmic parts of $f_1$ and $g_1$ in (\ref{f1:g1:expressions}):
$\hat{f}_1= 10.62$, $\hat{g}_1= 16.20$ for $\eta_c$ hadronic decay;
$\hat{f}_1= 9.73$, $\hat{g}_1= 15.06$ for $\eta_b$ decay.
Plugging these numbers into (\ref{parametrization:F1:G1}), one concludes
that the perturbation series in $F_1({}^1S_0)$ and $G_1({}^1S_0)$ in general have a poor convergence behavior,
which is particularly alarming
for $\eta_c$ decay due to the greater value of $\alpha_s$.

Substituting (\ref{f2:expressions}) into (\ref{parametrization:F1}), we then obtain the most
comprehensive formula for $\eta_c$ hadronic width within NRQCD factorization.
Demanding that the hadronic width (\ref{etac:decay:NRQCD:factorization}) is independent of
the factorization scale $\mu_\Lambda$, one readily deduce the following RGE:
\bqa
\frac{d\langle {\cal O}_1({}^1S_0)\rangle_{\eta_c}}{d\ln\mu_\Lambda^2} &=&
\alpha_s^2 \left(C_F^2+\frac{C_A C_F}{2}\right)
\langle {\cal O}_1({}^1S_0)\rangle_{\eta_c}
\nn\\
&-& \frac{4}{3}\frac{\alpha_s}{\pi} C_F \frac{\langle {\cal P}_1({}^1S_0)\rangle_{\eta_c}}{m^2} + \cdots.
\eqa
We have neglected the contribution from the operator ${\cal O}_8({}^1P_1)$,
which is suppressed by relative order-$v^4$~\cite{Bodwin:1994jh}.


Now we are ready to confront our state-of-the-art formula with the measured $\eta_{c,b}$ hadronic widths.
To facilitate the inclusion of the leading relativistic correction,
it is customary to introduce the following dimensionless ratio:
\beq
\langle v^2 \rangle_{\eta_c} ={\langle{\cal P}_1({}^1S_0)\rangle_{\eta_c} \over  m^2 \langle{\cal O}_1({}^1S_0)\rangle_{\eta_c}}.
\eeq
We adopt the following values of the encountered NRQCD matrix elements~\cite{Bodwin:2007fz,Chung:2010vz}:
\bqa
&& \langle{\cal O}_1({}^1S_0)\rangle_{\eta_c} = 0.470\,{\rm GeV^3}, \;\langle v^2 \rangle_{\eta_c} = \frac{0.430\,{\rm GeV^2}}{m_c^2},
\nn\\
&& \langle{\cal O}_1({}^1S_0)\rangle_{\eta_b} = 3.069\,{\rm GeV^3},   \;
\langle v^2 \rangle_{\eta_b} = -0.009.
\label{NRQCD:matrix:elements:etac:etab}
\eqa

\begin{figure}[tbh]
\includegraphics[width=0.45\textwidth]{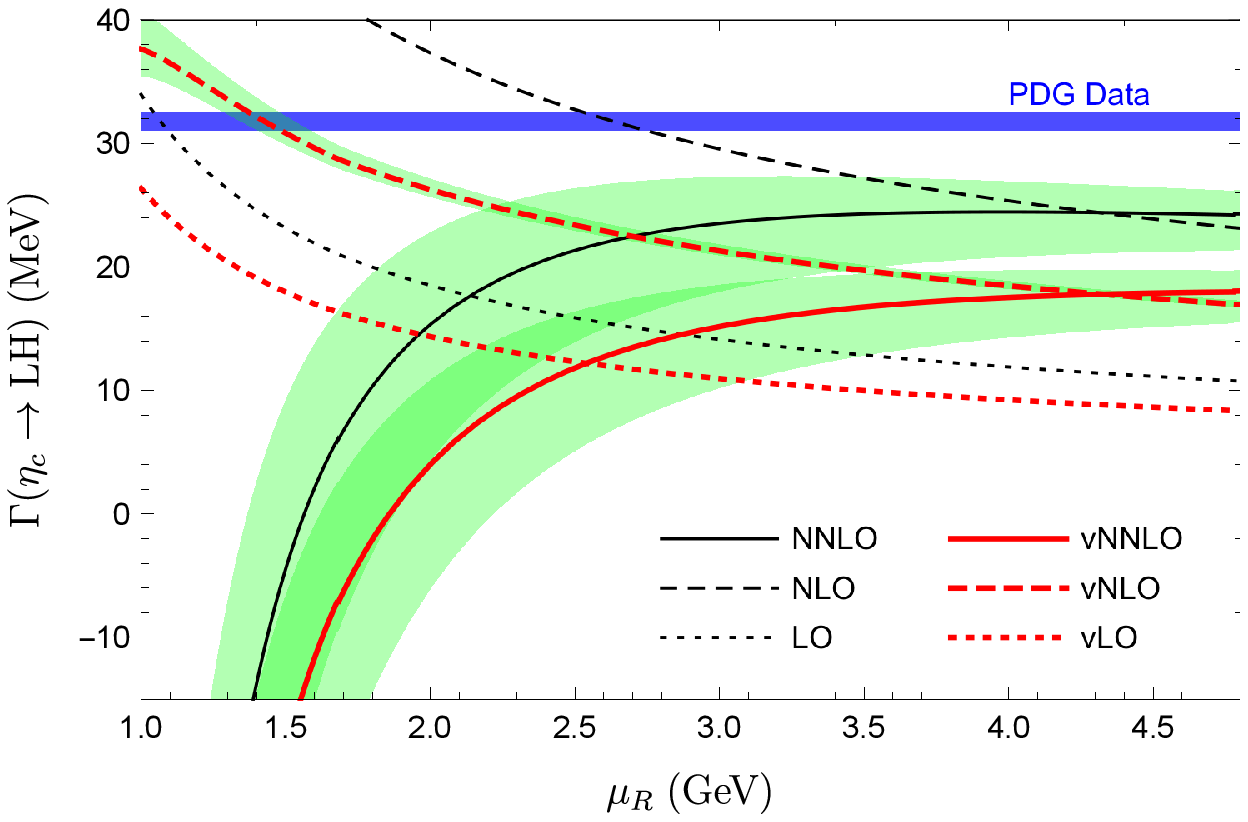}
\includegraphics[width=0.45\textwidth]{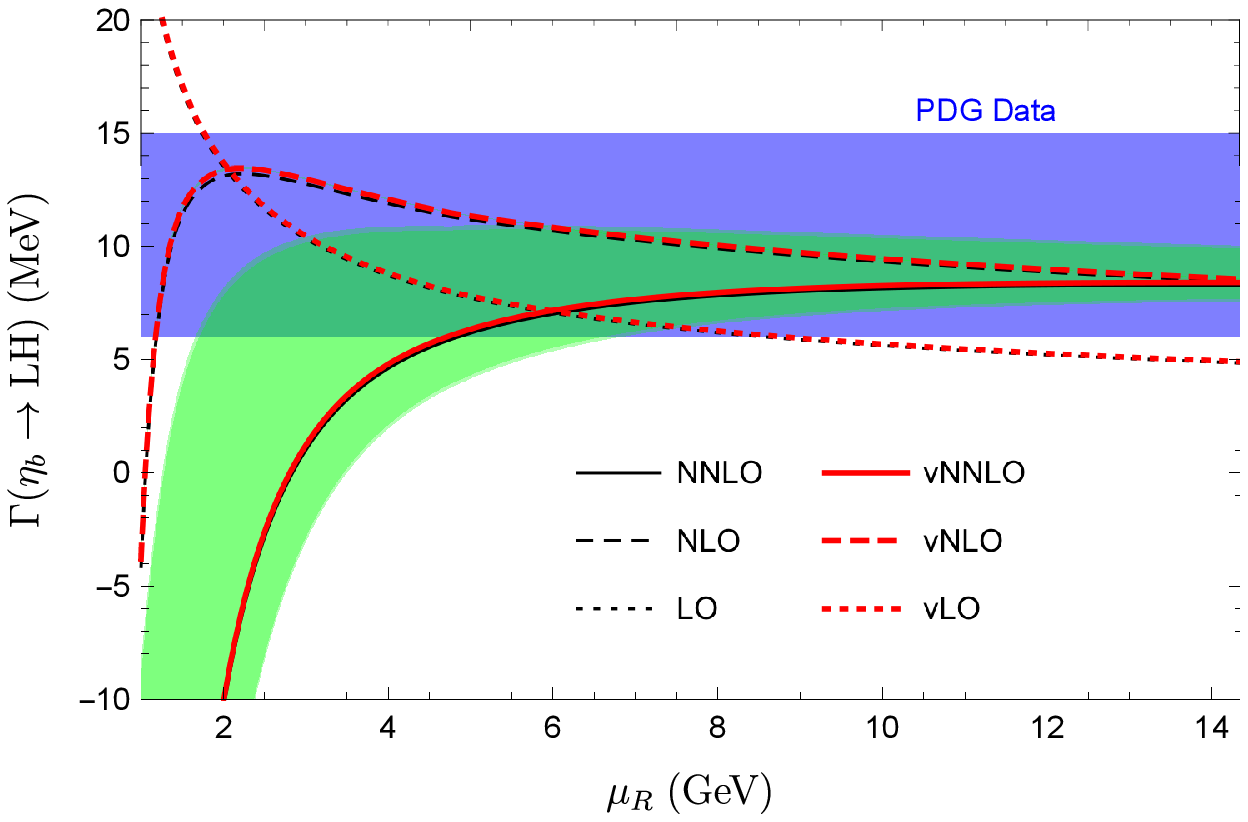}
\caption{\label{etacb:decay:widths}
The predicted hadronic widths of $\eta_c$ (top) and $\eta_b$ (bottom) as functions of $\mu_R$,
at various level of accuracy in $\alpha_s$ and $v$ expansion.
The blue bands correspond to the measured hadronic widths
$\Gamma_{\rm had}(\eta_c) = 31.8\pm0.8$ MeV, and $\Gamma_{\rm had}(\eta_b) = 10^{-4}_{+5}$ MeV~\cite{Olive:2016xmw}.
The label ``LO" represents  the NRQCD prediction at the lowest order in $\alpha_s$ and $v$,
and the label ``NLO" denotes the ``LO" prediction plus the ${\cal O}(\alpha_s)$ correction,
while  ``NNLO" signifies the ``NLO" prediction plus the ${\cal O}(\alpha_s^2)$
correction.
The label ``vLO" represents the ``LO" prediction together with the tree-level order-$v^2$ correction,
and ``vNLO" designates the ``vLO" prediction supplemented with the relative order-$\alpha_s$ and
order-$\alpha_s v^2$ correction, while ``vNNLO" refers to the ``vNLO" prediction further
supplemented with the order-$\alpha_s^2$ correction.
The green bands are obtained by varying $\mu_\Lambda$ from $1$ GeV to twice heavy quark mass,
and the central curve inside the bands are obtained by setting $\mu_\Lambda$ equal to heavy quark mass.}
\end{figure}

For phenomenological analysis, we take $m_c = 1.6\,{\rm GeV}$ and $m_b = 4.78\,{\rm GeV}$,
and use {\tt RunDec}~\cite{Chetyrkin:2000yt}
to compute the QCD running coupling at three-loop accuracy.

In Fig.~\ref{etacb:decay:widths}, we plot hadronic widths of $\eta_{c,b}$ as functions of $\mu_R$,
at various levels of accuracy in $\alpha_s$ and $v$ expansion.
For those nonperturbative matrix elements as chosen in (\ref{NRQCD:matrix:elements:etac:etab}),
we can observe some interesting patterns. For the hadronic $\eta_c$ width,
with the choice of lower renormalization scale, the LO and NLO predictions might be capable to account for the PDG data.
Nevertheless, with the inclusion of the NNLO perturbative correction, the NRQCD prediction in general becomes
significantly lower than the PDG data, even becomes negative for very small $\mu_R$.
A negative decay rate is certainly unphysical, which can be attributed to the large negative prefactor accompanying $\ln (4m^2/\mu_R^2)$ and the large negative non-logarithmic constant affiliated with the $\alpha_s^2/\pi^2$ in the decay rate.
Including relativistic corrections drives the prediction further away from the PDG data,
so that the discrepancy becomes even more pronounced.

For the hadronic $\eta_b$ width, the situation appears to be in a much better shape.
NRQCD prediction exhibits a quite satisfactory convergence behavior,
and our NNLO predictions are well compatible with the PDG measurements, albeit within large
experimental errors. The effects of relativistic corrections are too small to be discernible
in Fig.~\ref{etacb:decay:widths}.

A cautious reader may be skeptical about the objectiveness of our assertion,
due to the strong sensitivity of the predicted $\eta_c$ hadronic width to the input parameters
such as heavy quark mass and the NRQCD matrix elements. For this reason, next we turn to a much cleaner
experimental observable, the branching fraction of pseudoscalar quarkonium decay to two photons, ${\rm Br}(\eta_{c,b} \to \gamma\gamma)$, which is supposed to be much less contaminated by these nonperturbative factors.

After incorporating the known ${\cal O}(\alpha_s^2)$~\cite{Feng:2015uha,Czarnecki:2001zc}
and ${\cal O}(\alpha_s v^2)$~\cite{Jia:2011ah,Guo:2011tz} corrections,
we obtain the state-of-the-art predictions for the partial widths of $\eta_{c,b} \to \gamma\gamma$,
with an accuracy comparable to the hadronic widths of $\eta_{c,b}$ predicted in this Letter (Note some
alternative nonperturbative approaches have also predicted
the partial width of $\eta_c\to\gamma\gamma$~\cite{Dudek:2006ut,Chen:2016bpj}.).
Invoking the vacuum saturation approximation,
expanding the ratio in power series of $\alpha_s$ and $v$,
we find
\begin{subequations}
\bqa
&& {\rm Br}(\eta_c\to\gamma\gamma) = \frac{8\alpha^2}{9\alpha_s^2} \Bigg\{ 1 - \frac{\alpha_s}{\pi}\left[ 4.17\ln\frac{\mu_R^2}{4m_c^2} +14.00 \right] \nn\\
&&\quad\quad\quad\quad +\frac{\alpha_s^2}{\pi^2}\Bigg[4.34\ln^2\frac{\mu_R^2}{4m_c^2} +22.75\ln\frac{\mu_R^2}{4m_c^2} + 78.8 \Bigg] \nn\\
&&\quad\quad\quad\quad +2.24 \langle v^2\rangle_{\eta_c} \frac{\alpha_s}{\pi} \Bigg\},
\\
&& {\rm Br}(\eta_b\to\gamma\gamma) = \frac{\alpha^2}{18\alpha_s^2} \Bigg\{ 1 - \frac{\alpha_s}{\pi}\left[ 3.83\ln\frac{\mu_R^2}{4m_b^2} +13.11 \right] \nn\\
&&\quad\quad\quad\quad +\frac{\alpha_s^2}{\pi^2}\Bigg[3.67\ln^2\frac{\mu_R^2}{4m_b^2} +20.30\ln\frac{\mu_R^2}{4m_b^2} + 85.5 \Bigg] \nn\\
&&\quad\quad\quad\quad +1.91 \langle v^2\rangle_{\eta_b} \frac{\alpha_s}{\pi} \Bigg\}.
\eqa
\end{subequations}
Interestingly, not only the leading NRQCD matrix element $\langle{\cal O}_1({}^1S_0)\rangle_{\eta_c}$ cancels in the ratio, but also
the factorization scale $\mu_\Lambda$ cancels.
Note the branching fraction now depends on the heavy quark mass only logarithmically.

\begin{figure}[tbh]
	\centering
	\includegraphics[width=0.45\textwidth]{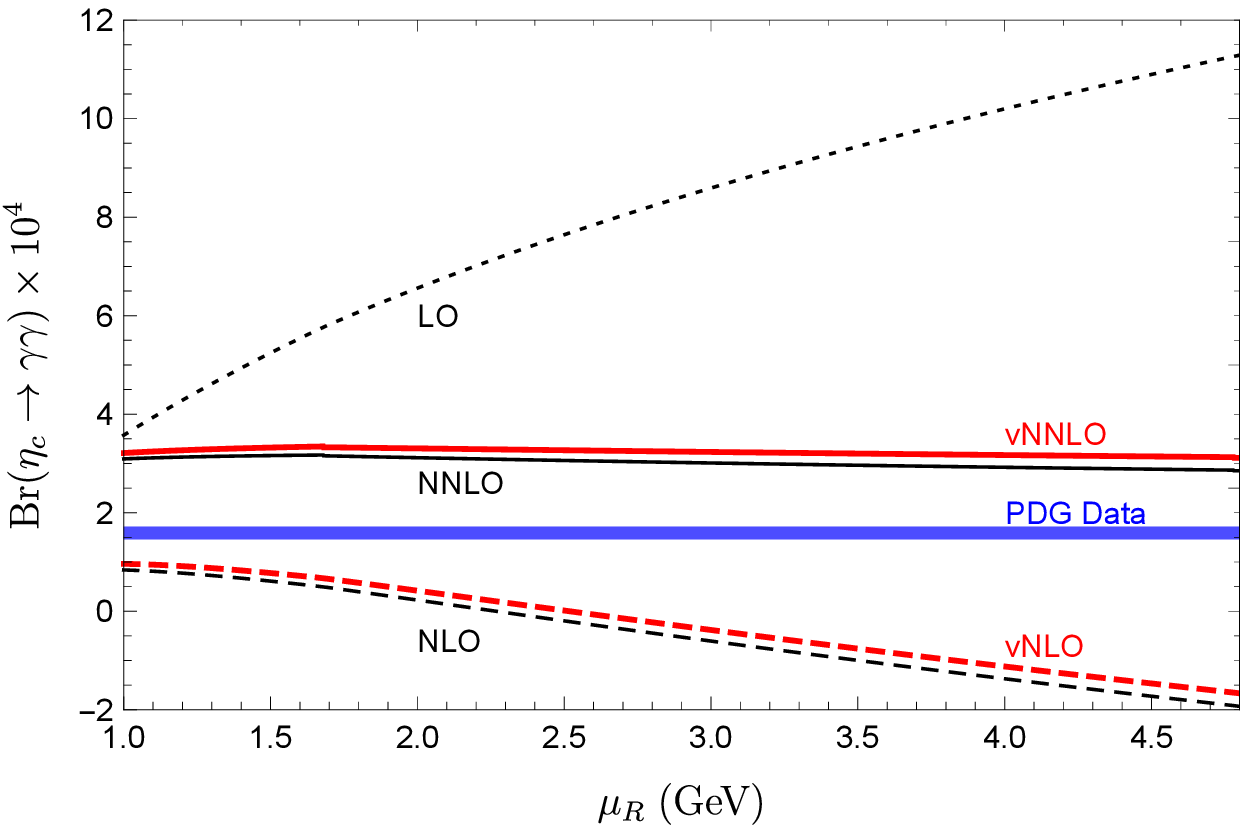}
	\includegraphics[width=0.45\textwidth]{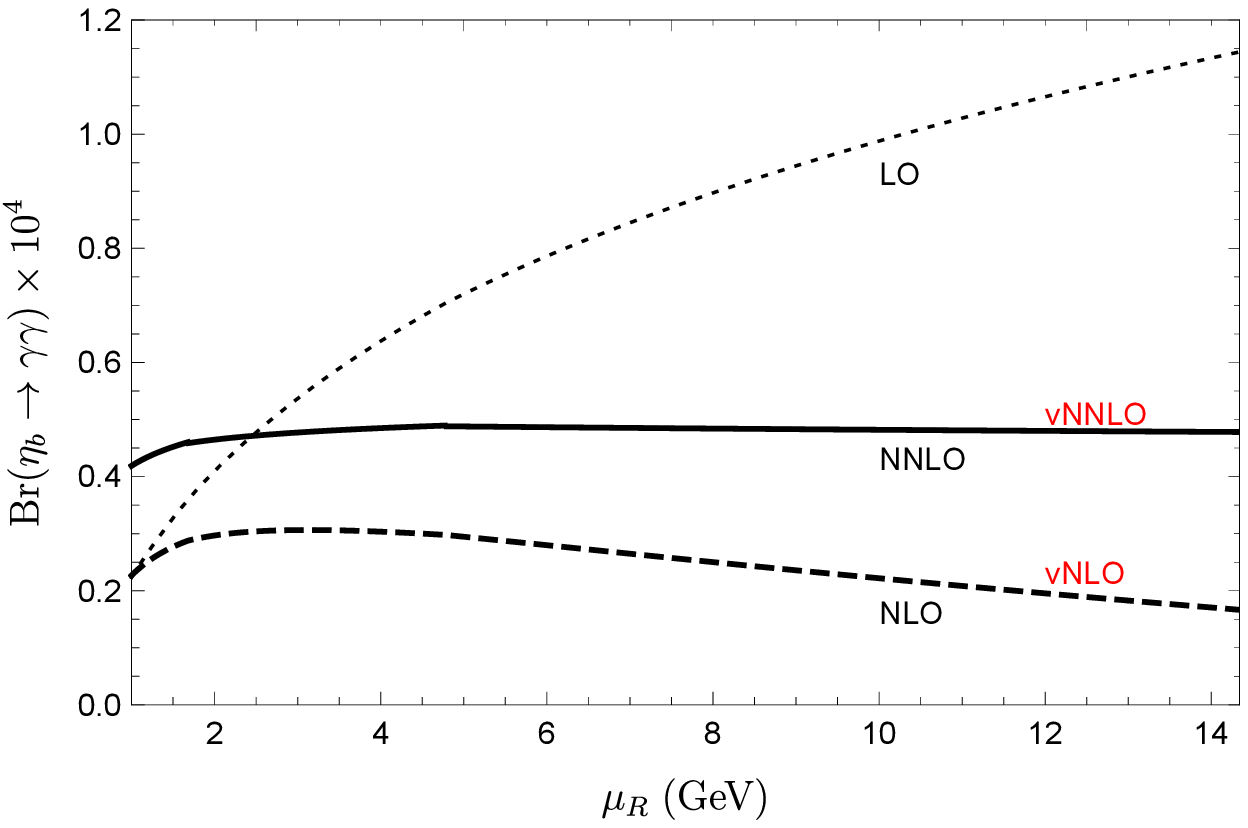}
	\caption{\label{Braching:Ratios:etacb:two:photons}
The predicted branching fractions of $\eta_c \to \gamma\gamma$ (top)
and $\eta_b \to \gamma\gamma$ (bottom) as functions of $\mu_R$, at various level of accuracy
in $\alpha_s$ and $v$. The blue band corresponds to
the measured branching ratio for $\eta_c\to\gamma\gamma$ taken from PDG~2016~\cite{Olive:2016xmw},
with ${\rm Br}(\eta_c\to\gamma\gamma) = (1.59 \pm 0.13) \times 10^{-4}$.
The labels characterizing different curves are the same as in Fig.~\ref{etacb:decay:widths}.}
\end{figure}

In Fig.~\ref{Braching:Ratios:etacb:two:photons}, we plot the NRQCD predictions to the branching fractions of $\eta_{c,b} \to \gamma\gamma$ as functions of $\mu_R$, at various levels of accuracy
in $\alpha_s$ and $v$. A curious feature is that the branching ratio exhibits much better convergence behavior than the hadronic width
itself. Moreover, the predicted branching ratio exhibits a very mild dependence on $\mu_R$,
even when $\mu_R$ gets small.
The relativistic correction also has rather mild effect.
Varying $\mu_R$ from 1 GeV to $3 m_c$, our state-of-the-art NRQCD predictions yield
${\rm Br}(\eta_c\to\gamma\gamma)$ ranging from
$3.1\times10^{-4}$ to $3.3\times10^{-4}$, which is more than $10\,\sigma$ away from the PDG value $(1.59 \pm 0.13) \times 10^{-4}$~\cite{Olive:2016xmw}! This sheer failure may indicate that, the NRQCD approach might
confront some serious troubles when applied to charmonium inclusive decay processes.

On the other hand, NRQCD approach appears to be much more trustworthy when applied to bottomonium decay.
As can be seen from Fig.~\ref{Braching:Ratios:etacb:two:photons}, varying $\mu_R$ from 1 GeV to $3m_b$,
we predict the branching fraction of $\eta_b\to\gamma\gamma$ through NNLO accuracy to be
\beq
{\rm Br}(\eta_b\to\gamma\gamma)= (4.8\pm 0.7)\times 10^{-5},
\eeq
with the caveat that the error estimate may be overly simpleminded.
It is exciting if the forthcoming \textsf{Belle~II} experiment can actually
observe this two-photon decay channel in the near future.

To summarize, in this Letter we have computed, for the first time, the NNLO perturbative corrections to
the hadronic widths of $\eta_{c,b}$, at the lowest order in $v$.
The validity of NRQCD factorization for inclusive quarkonium decay process has been explicitly verified
through relative order $\alpha_s^2$. As a byproduct, we are also able to infer the RGE for the leading NRQCD operator
${\cal O}_1({}^1S_0)$ through relative order-$\alpha_s^2$.
Incorporating this new ingredient of correction together with the existing relativistic corrections,
we have made a comprehensive study on the $\eta_{c,b}$ hadronic widths and the branching fractions for
$\eta_{c,b}\to \gamma\gamma$.  We find that severe tension arises between our state-of-the-art NRQCD predictions
and the measured $\eta_c$ hadronic width, and the tension in ${\rm Br}(\eta_c\to \gamma\gamma)$ is particularly
disquieting. In our opinion, this may signal a profound crisis for the influential NRQCD factorization approach --
whether it can be adequately applicable to charmonium decay or not. Our study supports the consensus that NRQCD should work for bottomonium decay decently well.
We have made a to date most refined prediction, ${\rm Br}(\eta_b\to \gamma\gamma) = (4.8\pm 0.7)\times 10^{-5}$,
which eagerly awaits the future experiments to conduct a critical examination.

\begin{acknowledgments}
{\noindent\it Acknowledgment.}
We thank Qing-Feng Sun for participating in the initial phase of this work.
We are grateful to Thomas Gehrmann for providing us with the analytic formula of one of the master integrals
in Ref.~\cite{Bonciani:2009nb}.
The work of F.~F. is supported by the National Natural Science Foundation of China under Grant No.~11505285,
and by the Fundamental Research Funds for the Central Universities.
The work of Y.~J. is supported in part by the National Natural Science Foundation of China under Grants No.~11475188,
No.~11261130311 (CRC110 by DGF and NSFC), by the IHEP Innovation Grant under contract number Y4545170Y2,
and by the State Key Lab for Electronics and Particle Detectors.
W.-L.~S. is supported by the National Natural Science Foundation of China under Grants No. 11447031 and
No. 11605144, and also
by the Fundamental Research Funds for the Central Universities under Grant No. XDJK2016C067.
This work is also supported by National Supercomputer Center in Guangzhou and
ScGrid/CNGrid.
The Feynman diagrams in this Letter are prepared by using \textsf{JaxoDraw}~\cite{Binosi:2003yf}.
\end{acknowledgments}

\end{document}